\documentclass[12pt]{article}
\usepackage{jheppub}
\usepackage{verbatim}
\usepackage{amsmath}
\usepackage{amssymb}
\usepackage{amsthm}
\usepackage{slashed}
\usepackage{amsfonts}
\usepackage[utf8]{inputenc}
\usepackage[normalem]{ulem}
\usepackage{tikz}
\usetikzlibrary{patterns}
\usepackage{pgfplots}
 \usepgfplotslibrary{fillbetween}

\def\be{\begin{equation}}
\def\ee{\end{equation}}
\def\ba#1\ea{\begin{align}#1\end{align}}
\def\bg#1\eg{\begin{gather}#1\end{gather}}
\def\bm#1\em{\begin{multline}#1\end{multline}}
\def\bmd#1\emd{\begin{multlined}#1\end{multlined}}

\def\({\left(}
\def\){\right)}
\def\[{\left[}
\def\]{\right]}
\def\<{\langle}
\def\>{\rangle}

\newcommand{\bfig}{\begin{figure}\begin{center}}
\newcommand{\efig}{\end{center}\end{figure}}
\newcommand{\bi}{\begin{itemize}}
\newcommand{\ei}{\end{itemize}}

\theoremstyle{definition}

\newcommand{\ZZ}{\mathbb{Z}}
\newcommand{\RR}{\mathbb{R}}

\newcommand{\rat}{\alpha}

\begin{document}

%\subheader{empty}
\title{Gravitating spinning strings in AdS$_3$}
\author[\flat]{Henry Maxfield}
\author[\sharp]{and Zhencheng Wang}
\affiliation[\flat]{Stanford Institute for Theoretical Physics, Stanford University, Stanford, CA 94305}
\affiliation[\sharp]{Department of Physics, University of California, Santa Barbara, CA 93106, USA}
\emailAdd{henrym@stanford.edu}
\emailAdd{zhencheng@ucsb.edu}

\abstract{In the AdS/CFT correspondence, single trace operators of large-$N$ gauge theories at large spin $J$ can be described by classical spinning strings, giving a geometric and classical description of their spectrum at strong coupling. We observe that in AdS$_3$ these strings have significant gravitational back-reaction at sufficiently large spin, since the gravitational force does not decay at long distances. We construct solutions for folded spinning strings coupled to gravity in AdS$_3$ and compute their spectrum, corresponding to the leading Regge trajectory of Virasroro primary operators. These solutions exist only below a maximal spin $J<J_\mathrm{max}$, and as $J\to J_\mathrm{max}$ the solution approaches an extremal rotating BTZ black hole.}

\maketitle

\section{Introduction}

It was suggested long ago that large-$N$ gauge theories might admit a description in terms of strings, beginning with 't Hooft's topological expansion of Feynman diagrams \cite{tHooft:1973alw}. This idea is realised concretely by the AdS/CFT correspondence \cite{Maldacena:1997re, Gubser:1998bc, Witten:1998qj}, with the surprising twist that the strings live in a higher dimensional spacetime with dynamical gravity. By studying the spectrum of states of a single string in AdS, we learn about the corresponding spectrum of single-trace operators in the dual CFT. Furthermore, at large spin $J\gg 1$ these strings become long in units of their tension, so that they can be described classically \cite{Gubser:2002tv}. Thus, we can learn about the spectrum of strongly-coupled gauge theories from the classical dynamics of spinning strings.

\begin{figure}
\centering
\begin{tikzpicture}
\draw [thick] (0,0) circle[radius=3];
\draw [fill] (0,0) circle[radius=0.05];
\draw [very thick,red] (-.925,-2.05) to (1.05,1.9) arc (-30:150:0.1) to (-1.05,-1.9) arc (150:300:0.1);
\draw [very thick, ->] (1.5,2) arc (50:75:2);
\node at (1.7,1.8) {$\omega$};
\end{tikzpicture}
\caption{A constant time slice of a folded classical string spinning around its center (the black dot) in AdS$_3$, with the black circle denoting the conformal boundary of AdS. Although we have drawn the two segments of the string separated for demonstrative reasons, they in fact coincide. The string spins like a rigid rod, with both ends spinning at the speed of light, moving along null geodesics.}
\label{fig:spinningstring}
\end{figure}
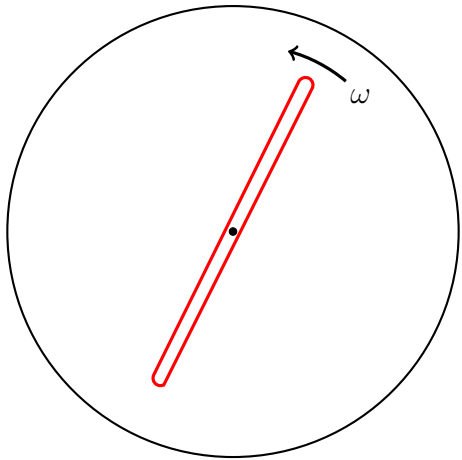

Of particular interest is the leading Regge trajectory, given by the single-trace operator of lowest conformal dimension $\Delta$ for each angular momentum $J$. The corresponding closed string state of lowest energy $E$ (equal to $\Delta$ in AdS units) for given $J$ is a folded string that rotates rigidly like a spinning rod, as sketched in figure \ref{fig:spinningstring}. For small $J$, this configuration has energy $E \propto \sqrt{T J}$ for string tension $T$, giving the famous linear Regge trajectories in the $J$-$E^2$ plane which are characteristic of strings in flat spacetime. At larger $J$  (of order $\ell^2 T$ for AdS length $\ell$), the spinning string lengthens enough to become sensitive to the curved geometry of AdS, modifying the spectrum. The linear Regge trajectories cross over to a logarithmically growing anomalous dimension $\Delta-J\sim 2\ell^2 T\log J$ for $J\gg \ell^2 T$ \cite{Gubser:2002tv}, a behaviour also seen in perturbative gauge theory \cite{Gross:1974cs,Georgi:1974wnj}.

In this paper, we observe that this is not the end of the story for strings in AdS$_3$ (times any compact internal space) and their dual two-dimensional CFTs. The reason is that gravitational backreaction becomes important for sufficiently large $J$. This phenomenon does not occur in higher dimensions simply because gravity decays at long distance, so the gravitational field sourced by a string does not increase as the string becomes longer. But in AdS$_3$, any source with total energy of order $G_N^{-1}$ (however diffuse it may be) has an order one effect on the metric even at infinite distance.% In addition, for such energies it becomes important to distinguish Virasoro primary states from states with additional boundary graviton excitations.

\begin{figure}
    \centering
    \includegraphics[width=\linewidth]{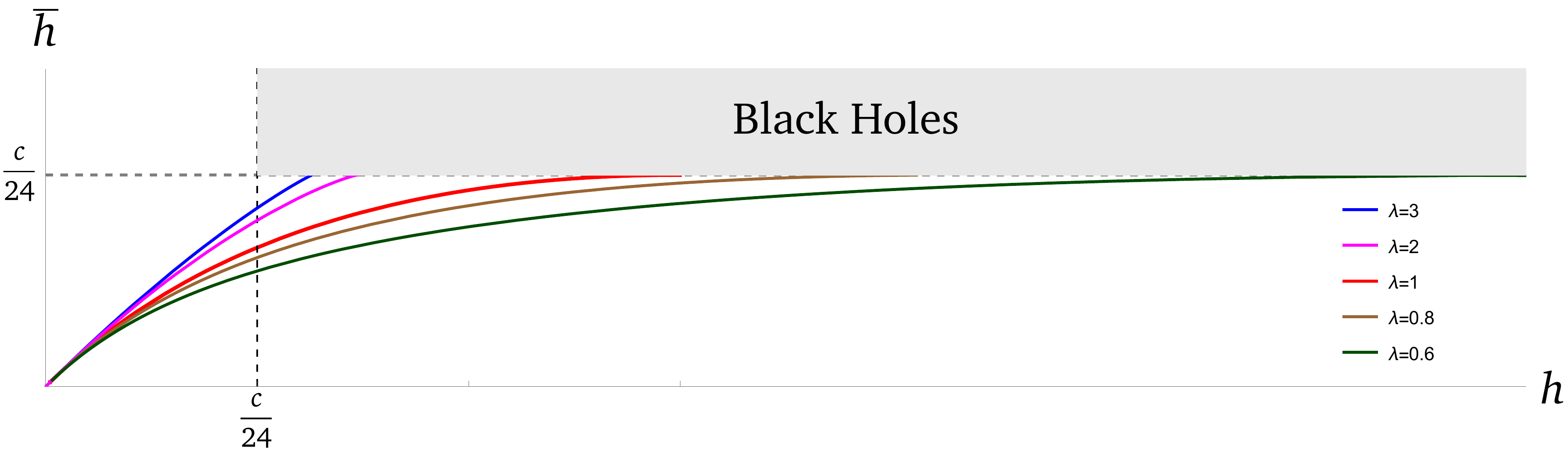}
    \caption{The string spectrum for various values of $\lambda$. We plot the relation between the conformal dimensions $h=\Delta+J$ and $\bar{h}=\Delta-J$ for the corresponding boundary states.  BTZ black holes cover the region where $h$ and $\bar{h}$ are both larger than $\frac{c}{24}$. For each $\lambda$, we have a one-parameter family of solutions running from zero energy and angular momentum to finite maximal values, where the string solution merges with an extremal black hole at $\bar h=\frac{c}{24}$.}
    \label{fig:stringspectrum}
\end{figure}

With this motivation, we will study classical solutions of a spinning Nambu-Goto string coupled to Einstein gravity with negative cosmological constant $\Lambda = -\frac{1}{\ell^2}$. This system is governed (classically) by a single dimensionless parameter $\lambda$ defined by
\begin{equation}
    \lambda = 8\pi \ell G_N T
\end{equation}
where $T$ is the string tension, alternatively expressed in terms of Regge slope or string length as $T=\frac{1}{2\pi\alpha'} = \frac{1}{2\pi\ell_s^2}$. We can think of $\lambda$ as the strength of gravitational coupling to a string with length of the same order as the AdS curvature scale $\ell$. With this simple theory, we will describe an ansatz for a folded closed spinning string and construct the one-parameter set of solutions for each $\lambda$ within this ansatz. The spectrum of these solutions --- the curve in the $h$-$\bar{h}$ plane described by their energy $\Delta=h+\bar{h}$ and angular momentum $J=h-\bar{h}$ --- delineates the leading Regge trajectory of lightest single-trace operators dual to such string states. These results are illustrated in figure \ref{fig:stringspectrum}.

In the case $\lambda\ll 1$ (which is perhaps most important for top-down models, see section \ref{sec:D1D5}), the regime of logarithmic anomalous dimensions mentioned above remains intact, but ultimately crosses over to a new behaviour. At first this is simply because the naive spinning strings are not Virasoro primary states (which classically correspond to spacetime-independent boundary energy-momentum tensor), but for larger spin still the strings source a large change in the geometry. For $\lambda$ of order unity, the logarthmic regime is absent, with back-reaction taking over before it is reached while the string size is of order the AdS length.

The most intriguing result is that these string solutions do not exist for arbitrarily large angular momentum. Instead, the family of solutions terminates at a maximal value of the angular momentum $J_\mathrm{max}$, where the spinning string smoothly becomes an extremal rotating BTZ black hole. As $J$ approaches $J_\mathrm{max}$ the geometry develops a long AdS$_2$ throat (as is familiar from extremal black holes), while the string which sources the geometry recedes deeper within that throat. This phenomenon is reminiscent of the black hole/string transition \cite{Susskind:1993ws,Bowick:1985af,Horowitz:1996nw,Giveon:2006pr,Chen:2021dsw}, a suggested correspondence between states of black holes and of single strings, though it differs in details from previous examples. Notably, in this case the proposed transition is to a black hole with large area (of order $\lambda^{-1}$ in AdS units for the most interesting case $\lambda\ll 1$), giving a large Bekenstein-Hawking entropy and more control over the solutions due to the classical limit. These string solutions provide great potential for more detailed study as an example of the black hole/string transition ideas.

% The regime of logarithmic anomalous dimensions described above corresponds to a string much longer than $\ell$, so survives only for $\lamda\ll 1$ so that back-reaction is negligible in this regime. This case of small $\lambda$ is perhaps most interesting (and expected in top-down constructions of CFTs with semiclassical gravity duals), but we will construct solutions for any value of $\lambda$.

There is a large literature on string solutions in AdS, much of it using integrability methods to construct a variety of string solutions of which the folded spinning string is the simplest, and relating to dual integrable spin chains: a small selection is \cite{Kruczenski:2004wg,Plefka:2005bk,Jevicki:2007aa,Jevicki:2009uz,Callebaut:2015fsa,Vegh:2015ska}. Some (such as \cite{David:2014qta,Banerjee:2015qeq}) make use of a special feature of AdS$_3$ which is absent in higher dimensions, namely a symmetric NS-NS $B$-field background, though we do not consider this here (see section \ref{sec:Bfield} for discussion). Perhaps the most similar previous work \cite{Kim:2014bga,Kim:2015bba} studied back-reacted solutions for circular strings (with rotational symmetry), carrying angular momentum only from internal excitations. Despite this large literature, the observation that folded strings in AdS$_3$ must always source significant back-reaction for large $J$ is novel.

The paper is organised as follows. The main work is developed in section \ref{sec:strings}, where we construct the string solutions of interest. We discuss the relevant equations of motion, describe the ansatz for folded spinning strings, provide solutions in terms of integrals, and finally give closed-form expressions for the energy and angular momentum of the solutions. In section \ref{sec:spectrum}, we analyse the spectrum in various limits of interest, as well as giving the exact spectrum in the case $\lambda=1$ which happens to enjoy technical simplifications. In section \ref{sec:extr} we give some details of the geometry and string solution as it approaches an extremal black hole. Finally, we discuss open questions, generalisations and speculations in section \ref{sec:discussion}.

\section{Spinning classical strings solutions in AdS$_3$ gravity}
\label{sec:strings}

\subsection{Strings coupled to gravity}

Our aim is to construct classical solutions of Einstein gravity in three dimensions with negative cosmological constant, sourced by a spinning string. Away from the string source, the spacetime satisfies the vacuum Einstein equations,
%\begin{equation}
%	S_\mathrm{EH} = \frac{1}{16\pi G_N}\int d^3 x\sqrt{-g}\left(\mathcal{R}+2\right)
%\end{equation}
\begin{equation}
	R_{\mu\nu} = -2g_{\mu\nu},
\end{equation}
where we have chosen units to set the AdS length $\ell$ to unity. In three dimensions, the Ricci tensor entirely determines the curvature, so the geometry is locally isometric to AdS. This greatly simplifies our analysis, because our solutions consist simply of a region of AdS bounded by the string worldsheet, with appropriate identifications.

The gravitational equations are sourced by the string, giving us a stress tensor localised on the worldsheet. The string dynamics is governed by the Nambu-Goto action (proportional to the area of the string worldsheet):
\begin{equation}
\label{eq:NGaction}
	S_{NG} = - T\int d\tau d\sigma  \, \sqrt{-\det h},
\end{equation}
where $\tau$, $\sigma$ are coordinates on the worldsheet and $h$ is the induced metric. The constant of proportionality is the string tension $T=\frac{1}{2\pi \alpha'}$.

Since the worldsheet is a timelike hypersurface in three dimensions, Einstein's equations at the worldsheet are equivalent to the Israel junction conditions \cite{Israel:1966rt} which relate the metric on either side of the worldsheet. First, the metric is continuous across the string, so the induced metric $h_{ab}$ of the string is the same determined from either side. Secondly, the metric has a discontinuous derivative at the string specified by a discontinuity of the extrinsic curvature of the worldsheet. For the Nambu-Goto string, this discontinuity is proportional to the induced metric:
\begin{equation}\label{eq:Kpm}
	K_{ab}^+ + K_{ab}^- = \lambda h_{ab},
\end{equation}
where 
\begin{equation}
	\lambda = 8\pi G_N T
\end{equation}
is a constant determining the strength of coupling between the string and gravity.
Here, $K_{ab}^\pm$ are the extrinsic curvature as determined from the metric on either side of the worldsheet. We define $K_{ab}^\pm$ with respect to an outward-pointing normal in both directions, which is why we have a sum of the two terms.

This gives us a complete set of equations of motion for the string coupled to gravity: the equations of motion from varying the string embedding are not independent. We can understand this by noting that varying the location of the string is equivalent to a variation of the metric by a diffeomorphism, while holding the coordinate location of the string fixed.

As a result, the general solution of strings coupled to AdS$_3$ gravity is given by a locally AdS$_3$ spacetime, obeying the junction conditions at the location of the worldsheet.

\subsection{Folded spinning string ansatz}

We will consider a particularly simple class of solutions, namely folded spinning strings invariant under a one-parameter continuous symmetry. We expect these solutions to correspond to the leading Regge trajectory of the string, that is the state of lowest energy for given angular momentum. This means that we have two coincident segments of string running between two points at which the string is folded back on itself; see figure \ref{fig:spinningstring}. This is equivalent to a spinning open string with massless endpoints (with twice the tension), so our solutions describe that case equally well.

The solutions we consider have a single continuous symmetry, a time translation along with a rotation at angular velocity $\omega$. We may write the corresponding Killing vector (which is tangent to the worldsheet) as $\kappa = \partial_t + \omega\partial_\phi$, where $\phi$ is an angular coordinate with period $2\pi$. We choose coordinates $(\tau,\sigma)$ on the string such that $\kappa = \partial_\tau$ on the worldsheet, and by imposing a conformal gauge so that the induced metric is given by
\begin{equation}\label{eq:confGauge}
	h = \Omega(\sigma)^2(-d\tau^2+d\sigma^2)
\end{equation}
for some positive function $\Omega(\sigma)$. This fixes our choice of $(\tau,\sigma)$ uniquely up to constant shifts; in particular, we may not separately specify the period of $\sigma$.

In addition to the continuous symmetry, we have two $\ZZ_2$ symmetries. First is a rotation by $\pi$, acting as $\phi\mapsto \phi+\pi$. We choose coordinates so that the `centre' of the spinning string, which is invariant under this rotation, lies at $\sigma=0$ (on one part of the folded string). The folded points of the string then lie at $\sigma = \pm \sigma_0$ for some $\sigma_0>0$, so that $\sigma$ is periodic with period $4\sigma_0$. The folds of the string $\sigma = \pm \sigma_0$ will move at the speed of light, so the conformal factor vanishes there: $\Omega(\sigma_0)=0$. The rotation symmetry acts on the worldsheet as $(\tau,\sigma)\mapsto (\tau,\sigma+2\sigma_0)$

Our second $\ZZ_2$ symmetry is a simultaneous reflection in space and time. In particular, this fixes the $\tau=0$ slice of the string while exchanging the two sides of the worldsheet. This symmetry relates the extrinsic curvatures $K^\pm$ on either side of the string, as
\begin{equation}
	K^+_{\tau\tau} = K^-_{\tau\tau},\quad K^+_{\sigma\sigma} = K^-_{\sigma\sigma}, \quad K^+_{\tau\sigma} = -K^-_{\sigma\tau}. 
\end{equation}
With this, junction conditions \eqref{eq:Kpm} can be written in terms of the extrinsic curvature on just one side of the string,
\begin{equation}\label{eq:stringEq}
	K_{\tau\tau}= -\lambda \Omega^2,\qquad K_{\sigma\sigma} = \lambda \Omega^2,
\end{equation}
while the off-diagonal curvature $K_{\tau\sigma}$ is unconstrained. Note that we have an additional factor of two in the junction conditions, since we have two coincident strands of string.

Next we choose coordinates for AdS$_3$ and specify the string embedding. Outside the string, we can write the most general locally AdS$_3$ metric in Fefferman-Graham coordinates, defined for us with the conformal boundary located at $z\to\infty$ (comparing to some other common conventions, $z$ is the inverse of the radial Fefferman-Graham coordinate or the square of its inverse). This metric is
\begin{equation}\label{eq:gFG}
	ds^2 = \tfrac{1}{4} \epsilon_L du^2 + \tfrac{1}{4} \epsilon_R dv^2 -\left(z+\frac{\epsilon_L\epsilon_R}{16z}\right)dudv + \frac{dz^2}{4z^2},
\end{equation}
where $u,v$ become lightcone coordinates on the boundary $z\to\infty$:
\begin{equation}\label{eq:uv}
	u = t+\phi,\quad v= t-\phi, \qquad (u,v)\sim (u+2\pi,v-2\pi),
\end{equation}
where the identification of coordinates comes from the $2\pi$ periodicity of the angle $\phi$. While these coordinates are convenient for calculations, the metric perhaps looks more familiar if we substitute $z$ for a radial coordinate $r$ defined by $r^2 = \frac{1}{z}(z+\frac{\epsilon_L}{4})(z+\frac{\epsilon_R}{4})$ (the coefficient of $d\phi^2$), giving
\begin{equation}\label{eq:metricr}
\begin{gathered}
      ds^2 = -f(r) dt^2 +\frac{dr^2}{f(r)} + r^2\left(d\phi-\frac{\epsilon_R-\epsilon_L}{4r^2}dt\right)^2, \\ \text{where } f(r) = r^2 - \frac{\epsilon_R+\epsilon_L}{2} + \frac{(\epsilon_R-\epsilon_L)^2}{16r^2} \,.
\end{gathered}
\end{equation}

The coefficients $\epsilon_L,\epsilon_R$ appearing in the metric are proportional to the left- and right-moving energies,
\begin{equation}
	E_L = \frac{\epsilon_L}{16G_N}, \quad E_R = \frac{\epsilon_R}{16G_N}
\end{equation}
with $E=E_R+E_L$ giving the energy and $J=E_R-E_L$ the angular momentum of the solution. In particular, the global AdS$_3$ vacuum corresponds to $\epsilon_L=\epsilon_R=-1$, with negative Casimir energy $E_\mathrm{vac} = -\frac{1}{8G_N}$. With these solutions and boundary metric $ds^2=-dudv=-dt^2+d\phi^2$, the stress tensor is independent of $t$ and $\phi$.

This constant stress tensor corresponds to considering Virasoro primary states of the string: the stress tensor Fourier modes in $u$ and $v$ are the Virasoro generators $L_n$, $\bar{L}_n$, and in a primary state their expectation values vanish except for the constant modes $n= 0$. Since a classical state has small fluctuations, constant expectation value suffices to guarantee that the state has $E_L$ and $E_R$ close to that of a Virasoro primary. See section 5.2 of \cite{Collier:2018exn} for a similar discussion in the context of two-particle states.

More general states can be constructed from the same metric by performing a conformal transformation (separate diffeomorphisms on $u$ and $v$ coordinates, followed by a Weyl transormation to a flat metric in the new coordinates), corresponding to a particular class of `coherent' Virasoro descendants of the original state. 

We may also express these parameters in terms of standard CFT variables, using the Brown-Hennaux relation $c\sim \frac{3}{2G_N}$ \cite{Brown:1986nw}. The states we construct correspond to Virasoro primary operators of conformal dimension $\Delta$ and spin $J$, with
\begin{equation}
\begin{gathered}
	h = \frac{c}{24}(\epsilon_R+1),\qquad \bar{h} = \frac{c}{24}(\epsilon_L+1),\\
	\text{where }\qquad\Delta = h+\bar{h},\qquad J = h-\bar{h} \;.\qquad
\end{gathered}
\end{equation}

In these coordinates, the symmetry of our solution acts as a translation in the $u$ and $v$ coordinates:
\begin{equation}
	\kappa = (\omega+1)\partial_u - (\omega-1)\partial_v \,.
\end{equation}
With this, we can write the string embedding in terms of unknown functions $u_0$, $v_0$, $z_0$ depending on $\sigma$ only, specifying the string at $\tau=0$:
\begin{equation}
	u(\tau,\sigma) = u_0(\sigma) + (\omega+1) \tau,\quad v(\tau,\sigma) = v_0(\sigma) - (\omega-1) \tau, \quad z(\tau,\sigma) = z_0(\sigma).
\end{equation}

%At this point we choose coordinates for AdS$_3$ to specify the embedding of the string and compute the extrinsic curvatures. To make things as simple as possible, we would like the symmetry $\kappa$ to act as a translation. \HM{Motivate coords...}
% As we will see, this in fact requires two distinct choices depending on the energy and angular momenumtum, which govern the conjugacy class of $\kappa$ in the $\mathfrak{sl}(2,\RR)\oplus\mathfrak{sl}(2,\RR)$ symmetry algebra of AdS$_3$.
%It is convenient to use lightcone-like coordinates $u,v,r$
%\begin{equation}
%	ds^2 = -\tfrac{1}{4}(du+dv)^2 -r^2 du dv + \frac{dr^2}{1+r^2}
%\end{equation}
%
%\begin{equation}
%	(u,v)\sim (u+\Delta u,v\sim v - \Delta v)
%\end{equation}

%\begin{equation}
%	t= \pi\frac{u}{\Delta u}+\pi\frac{v}{\Delta v}, \quad \phi = \pi\frac{u}{\Delta u}-\pi\frac{v}{\Delta v}
%\end{equation}
%so $\phi$ has period $2\pi$.

%\begin{equation}
%E=-\frac{1}{16G_N} \frac{(\Delta u)^2+(\Delta v)^2}{(2\pi)^2}
%\end{equation} 
%\begin{equation}
%J=\frac{1}{16G_N} \frac{(\Delta u)^2-(\Delta v)^2}{(2\pi)^2}.
%\end{equation} 

With this ansatz, we can outline our general strategy for constructing solutions:
\begin{itemize}
	\item Impose conformal gauge \eqref{eq:confGauge}. This fixes $u_0'(\sigma)$ and $v_0'(\sigma)$ in terms of $z_0(\sigma)$.
	\item Compute the extrinsic curvature of the string embedding in terms of the parameters $\omega,\epsilon_L,\epsilon_R$ and the unknown function $z_0(\sigma)$.
	\item Use the first equation of \eqref{eq:stringEq} (relating the $\tau\tau$ components of extrinsic curvature and induced metric on the worldsheet) to solve for $z_0(\sigma)$ in terms of $\lambda,\omega,\epsilon_L,\epsilon_R$. When this is imposed, the second equation ($\sigma\sigma$ component) turns out to be automatically satisfied.
	\item Using this solution for $z_0$, impose the correct periodicity for $u_0$, $v_0$ from \eqref{eq:uv} to fix $\epsilon_L$, $\epsilon_R$ in terms of $\lambda$, $\omega$.
\end{itemize}
For given string tension parameter $\lambda$, we will thus construct a one-parameter family of solutions, corresponding to the leading single-trace Regge trajectory of Virasoro primary states.

Before describing the detailed implementation of this strategy, we briefly comment on the interpretation of the metric \eqref{eq:gFG} outside the string as a quotient of AdS$_3$. The isometry algebra of AdS$_3$ is $\mathfrak{sl}(2,\RR)_L \oplus \mathfrak{sl}(2,\RR)_R$, with elements given by pairs $(\xi_L,\xi_R)$ of elements of $\mathfrak{sl}(2,\RR)$. Our metric outside the string is a quotient by the group generated by a single finite isometry $(g_L,g_R)=(\exp\xi_L,\exp\xi_R)$. We have chosen coordinates such that the commuting symmetries $\xi_L$ and $\xi_R$ both act as translations of coordinates $u,v$ respectively. The parameters $\epsilon_{L,R}$ are then determined by the conjugacy classes of $\xi_{L,R}$ in $\mathfrak{sl}(2,\RR)$. In particular, the sign of $\epsilon$ depends on whether the corresponding $\xi$ is in an  elliptic ($\epsilon<0$), parabolic ($\epsilon=0$), or hyperbolic ($\epsilon>0$) conjugacy class. The familiar BTZ solution corresponds to cases $\epsilon_L>0$ and $\epsilon_R>0$; extremal rotating BTZ is given by $\epsilon_L=0$ and $\epsilon_R>0$ (or vice-versa).

Finally before constructing our solutions, we note that if $\epsilon_{L}<0$ then $\partial_\phi$ becomes timelike in the region $z<-\frac{1}{4}\epsilon_{L}$. This means that the string cannot lie entirely in that region, since otherwise a circle of constant $t,z$ (with $z$ large enough to stay outside the string) would be a closed timelike curve.

\subsection{The string solutions}

The conformal factor for the induced metric on the string \eqref{eq:confGauge} can be read off simply from the $\tau\tau$ component, or equivalently the norm of the Killing field $\kappa$ (since $\kappa=\partial_\tau$ on the worldsheet):
\begin{equation}\label{eq:Omegaz}
	\Omega^2 = (\omega^2-1)\frac{(z_L-z_0)(z_0-z_R)}{z_0}\; ,
\end{equation}
where we have defined parameters giving the zeroes of $\Omega$ by
\begin{equation}
	z_L = -\frac{1}{4}\frac{\omega+1}{\omega-1}\epsilon_L,\qquad z_R = -\frac{1}{4}\frac{\omega-1}{\omega+1}\epsilon_R\,.
\end{equation}
The outermost points of the string where it folds follow a null trajectory, so are located at one of these zeroes, which without loss of generality we may choose to be $z_L$. For physical spinning string solutions, this will be the larger root (so $z_L>z_R$). We will always have $\omega>1$ so that $\Omega>0$. In particular, since $z_L>0$,  we will always have $\epsilon_L<0$. Note also that $z_L>-\frac{1}{4}\epsilon_L$, so there are no closed timelike curves of constant $z$.

These solutions all have positive angular momentum, $J>0$: oppositely-spinning  solutions with  $J<0$ and $\omega<-1$ are obtained by instead choosing the folds to be at $z=z_R$.

Next, by fixing the remaining components of the induced metric to conformal gauge, we find
\begin{align}
	u_0' &= (\omega+1) \frac{z_0^2-2z_R z_0+z_Lz_R}{z_0^2-z_Lz_R} \sqrt{1-\frac{1}{4\Omega^2}\frac{(z_0')^2}{z_0^2}} \;, \label{eq:u0p}\\
	v_0' &= (\omega-1) \frac{z_0^2-2z_L z_0+z_Lz_R}{z_0^2-z_Lz_R} \sqrt{1-\frac{1}{4\Omega^2}\frac{(z_0')^2}{z_0^2}} \, ,\label{eq:v0p}
\end{align}
where $'$ denotes derivative with respect to $\sigma$ throughout.

Using this, we can now compute the extrinsic curvature in terms of $z_0(\sigma)$. The important component for us (and the simplest to compute) is
\begin{equation}
	K_{\tau\tau} = (\omega^2-1) \frac{z_0^2-z_Lz_R}{z_0}\sqrt{1-\frac{1}{4\Omega^2}\frac{(z_0')^2}{z_0^2}} \,.
\end{equation}
Equating this with $\lambda\Omega^2$ as in  \eqref{eq:stringEq}, we find
\begin{equation}\label{eq:z0p}
	(z_0')^2 = 4(\omega^2-1)z_0(z_L-z_0)(z_0-z_R)\left[1-\left(\lambda\frac{(z_L-z_0)(z_0-z_R)}{z_0^2-z_Lz_R}\right)^2\right].
\end{equation}
Solving this sepearable ODE gives us the function $z_0(\sigma)$, and then \eqref{eq:u0p}, \eqref{eq:v0p} give $u_0(\sigma)$, $v_0(\sigma)$ as integrals to determine the string embedding. The centre of the string $\sigma=0$ lies at the value of $z$ where the contents of the square brackets vanishes, so $z_0'(0)=0$.

 The additional equation $K_{\sigma\sigma} = \lambda \Omega^2$ carries no further information: it is automatically satisfied as long as \eqref{eq:u0p}, \eqref{eq:v0p} and \eqref{eq:z0p} hold.
 
 \subsection{The string energy and angular momentum}

Substituting \eqref{eq:z0p} in \eqref{eq:u0p} and \eqref{eq:v0p} we find the following:
\begin{align}
	u_0' &= \lambda(\omega+1) \frac{(z_0^2-2z_Rz_0+z_Lz_R)(z_L-z_0)(z_0-z_R)}{(z_0^2-z_Lz_R)^2}   \label{eq:u0p2}\\
	v_0' &= \lambda(\omega-1) \frac{(z_0^2-2z_Lz_0+z_Lz_R)(z_L-z_0)(z_0-z_R)}{(z_0^2-z_Lz_R)^2}.\label{eq:v0p2}
\end{align}

We have one final condition to impose, namely the correct periodicity for the coordinates. This means that $u_0$ and $-v_0$ increase by $2\pi$ on going round the string, so they increase by $\frac{\pi}{2}$ between the centre and a fold. We can impose this by integrating the expressions for $u_0',v_0'$: $\int_0^{\sigma_0} u_0'(\sigma)d\sigma = -\int_0^{\sigma_0} v_0'(\sigma)d\sigma = \frac{\pi}{2}$.

At first this looks like it will give rise to a complicated implicit equation relating the four paramaters $z_L,z_R,\omega$ and $\lambda$ on which the solutions depend. In fact, it is rather simpler than it seems, since there are only really two parameters on which the solutions depend nontrivially, namely $\lambda$ and the ratio between $z_L$ and $z_R$, which we define as
\begin{equation}\label{eq:defr}
	\rat^2 = \frac{z_R}{z_L} = \left(\frac{\omega-1}{\omega+1}\right)^2\frac{\epsilon_R}{\epsilon_L}.
\end{equation}
We have $\rat^2<1$ (since $z_L>z_R$), but note that $\rat^2$ may be negative, so $\rat$ can be pure imaginary. The other parameters only provide simple proportionality constants. We will find explicit expressions for $\epsilon_{L,R}$ which depend only on $\lambda,\rat$ (a one-parameter family of solutions for a given string tension), and using this one may recover the value of $\omega$ from \eqref{eq:defr}.

This can be understood from the metric \eqref{eq:gFG} we started with. We can freely rescale $\epsilon_L$, $\epsilon_R$ by positive factors by absorbing them into a rescaling of the coordinates $u,v,z$. Using this freedom, one may set the metric to some fiducial choice which does not depend on any free parameters (only discretely on the signs of $\epsilon_{L,R}$), at the expense of changing the periodicity of $u,v$ by factors of $\sqrt{|\epsilon_{L,R}|}$. In this rescaled metric, the string solutions explicitly depend only on $\lambda$ and $\rat$ (which determines the solution's symmetry), and we ultimately recover $\epsilon_{L,R}$ from reading off the necessary scaling of $u,v$ to impose the appropriate periodicity.

The simplest expressions for $\epsilon_{L,R}$ are obtained by passing to a new radial coordinate $x$ on the string proportional to the conformal factor,
\begin{equation}\label{eq:xcoord}
	x = \frac{1}{\sqrt{z_L(\omega^2-1)}} \Omega=\sqrt{\frac{(z_L-z_0)(z_0-z_R)}{z_L z_0}},
\end{equation}
where equation \eqref{eq:Omegaz} is used. 
In terms of this new coordinate, the equation \eqref{eq:z0p} for the string embedding $x(\sigma)$ becomes
\begin{equation}
	(x')^2 = z_L(\omega^2-1)q(x),
\end{equation}
where $q$ is the quartic
\begin{equation}\label{eq:q}
	q(x) = (1-\lambda^2)x^4-2(1+\rat^2)x^2 + (1-\rat^2)^2\,.
\end{equation}
From this, one can solve for $x(\sigma)$ in terms of elliptic functions. Moreover, the expressions \eqref{eq:u0p}, \eqref{eq:v0p} become simpler in terms of this coordinate:
\begin{align}
	u_0' &= \lambda\frac{\omega+1}{2}\left[\frac{(1+\rat)^3}{(1+\rat)^2-x^2}+\frac{(1-\rat)^3}{(1-\rat)^2-x^2}-2\right], \\
	v_0' &= \lambda\frac{\omega-1}{2\rat}\left[\frac{(1+\rat)^3}{(1+\rat)^2-x^2}-\frac{(1-\rat)^3}{(1-\rat)^2-x^2}-2\rat\right].
\end{align}
With this, we can impose periodicity of $u,v$ as the integrals
\begin{equation}
	\frac{\pi}{2}= \int_0^{x_0} \frac{u_0'}{x'} dx,\qquad -\frac{\pi}{2}= \int_0^{x_0} \frac{v_0'}{x'} dx \; ,
\end{equation}
with limits $x=0$ corresponding to a fold of the string ($\sigma=\sigma_0$), and $x=x_0$ given by the smallest positive root of $q$ corresponding to the centre. For $\lambda<1$, requiring existence of such a root means that $\rat^2$ cannot be too negative.\footnote{Explicitly, $x_0^2=\frac{-1-\rat^2+\sqrt{\lambda^2 \rat^4+(4-2\lambda^2)\rat^2+\lambda^2}}{\lambda^2-1}$, reality requiring $\rat^2>\frac{-2+\lambda^{2}+2 \sqrt{1-\lambda^{2}}}{\lambda^{2}}$ when $\lambda<1$.\label{foot:alphamin}} From this, we obtain integral expressions for $\epsilon_{L,R}$:
\begin{align}
	\sqrt{-\epsilon_L}  &= \lambda\frac{2}{\pi}\int_0^{x_0}\frac{dx}{\sqrt{q(x)}} \left[\frac{(1+\rat)^3}{(1+\rat)^2-x^2}+\frac{(1-\rat)^3}{(1-\rat)^2-x^2}-2\right], \label{eq:epsLint} \\
	\sqrt{-\epsilon_R} &= -\lambda\frac{2}{\pi}\int_0^{x_0}\frac{dx}{\sqrt{q(x)}}\left[\frac{(1+\rat)^3}{(1+\rat)^2-x^2}-\frac{(1-\rat)^3}{(1-\rat)^2-x^2}-2\rat\right].  \label{eq:epsRint}
\end{align}
Note that for $z_R<0$, $\rat$ will be pure imaginary. In that case, the first two terms in the integrand of \eqref{eq:epsLint} add up to the real part of $\frac{(1+\rat)^3}{(1+\rat)^2-x^2}$ so we get a real result and $\epsilon_L<0$; in \eqref{eq:epsRint} they give $i$ times the imaginary part of $\frac{(1+\rat)^3}{(1+\rat)^2-x^2}$ so the integral is imaginary and $\epsilon_R>0$.

These integrals can be evaluated as
\begin{align}
	\sqrt{-\epsilon_L} &= \lambda\frac{2}{\pi}\left[\tfrac{x_0}{1-\rat}\Pi\left(\left(\tfrac{x_0}{1+\rat}\right)^2\big|m\right) +\tfrac{x_0}{1+\rat}\Pi\left(\left(\tfrac{x_0}{1-\rat}\right)^2\big|m\right) - \frac{2x_0}{1-\rat^2}K(m) \right] \label{eq:rootepsL} \\
	\sqrt{-\epsilon_R} &= -\lambda\frac{2}{\pi}\left[\tfrac{x_0}{1-\rat}\Pi\left(\left(\tfrac{x_0}{1+\rat}\right)^2\big|m\right) -\tfrac{x_0}{1+\rat}\Pi\left(\left(\tfrac{x_0}{1-\rat}\right)^2\big|m\right) - \frac{2\rat x_0}{1-\rat^2}K(m) \right] \label{eq:rootepsR}
\end{align}
where $K$ and $\Pi$ are complete elliptic integrals of the first and third kind\footnote{These are defined by $K(m)=\int_0^1 \frac{dt}{\sqrt{(1-t^2)(1-mt^2)}}$ and $\Pi (n|m)=\int_0^1 \frac{dt}{(1-nt^2)\sqrt{(1-t^2)(1-mt^2)}}$.} respectively, and
\begin{equation}
	m = x_0^4 \frac{1-\lambda^2}{(1-\rat^2)^2}<1
\end{equation}
is a ratio of roots of $q$ (so that $q(x)=m^{-1}(1-\lambda^2)(x_0^2-x^2)(x_0^2-m x^2)$).

By replacing these complete elliptic integrals with incomplete elliptic integrals as a function of $\frac{x}{x_0}$, one can determine the embedding of the string giving $u_0,v_0$ as a function of $x$.

Note that consistency of this solution requires that the expressions in \eqref{eq:epsLint} or \eqref{eq:rootepsL} give a positive value for $\sqrt{-\epsilon_L}$ (a negative value would demand a periodic identification of time rather than space).

\section{The string spectrum}
\label{sec:spectrum}

The solutions we have obtained and the expressions for $\epsilon_{L,R}$ determining their spectrum are a little too complicated for their properties to be manifest. In this section we explore various limits and special cases to illustrate various interesting features.

\subsection{Flat space limit}

We first study the limit where the string is small compared to the AdS scale, to check that we recover the physics of strings in flat space (backreaction is unimportant in this regime). This is the limit $\rat\to 1$ with $\lambda$ held fixed.

% $\epsilon_{L,R}\to -1$, $r\to1$, $\omega\to\infty$, $1-r\sim\frac{2}{\omega}$.
 The parameters appearing in \eqref{eq:rootepsL}, \eqref{eq:rootepsR} scale as
 \begin{equation}
 	x_0 = (1-\rat) - \frac{\lambda^2}{8}(1-\rat)^3 + O((1-\rat)^4), \quad m= \frac{1-\lambda^2}{4}(1-\rat)^2 +  O((1-\rat)^{3}).
 \end{equation}
From this, we need only make use of the limiting behaviour of the elliptic functions in various limits.

For the first terms in \eqref{eq:rootepsL}, \eqref{eq:rootepsR}, we use the series expansion at small $n,m$,
\begin{equation}
	\Pi(n|m) \sim \frac{\pi}{2}\left(1+\frac{m}{4} + \frac{n}{2} + \cdots \right) \qquad (n,m\to 0),
\end{equation}
to find
\begin{equation}
	\tfrac{x_0}{1-\rat}\Pi\left(\left(\tfrac{x_0}{1+\rat}\right)^2\big|m\right) \sim \frac{\pi}{2} +\frac{3\pi}{32}(1-\lambda^2)(1-\rat)^2 + \cdots.
\end{equation}
For the second terms, we instead require
\begin{equation}
	\Pi(n|m) \sim \frac{\pi}{2\sqrt{1-n}}\left(1+\frac{m}{2} + O(m^2) \right) +O(m,\sqrt{1-n}) \quad (n\to1, \; m\to 0),
\end{equation}
which gives
\begin{equation}
	\tfrac{x_0}{1+\rat}\Pi\left(\left(\tfrac{x_0}{1-\rat}\right)^2\big|m\right) \sim \frac{\pi}{2\lambda} + O((1-\rat)^3).
\end{equation}
The last terms are simple to expand using $K(m) = \frac{\pi}{2} +\frac{\pi}{8}m+O(m^2)$.
%\begin{equation}
%	\frac{2rx_0}{1-r^2}K(m) \sim \frac{\pi}{2} + \frac{\pi}{4}(1-r) + \frac{\pi}{32}(5-3\lambda^2)(1-r)^2+\cdots
%\end{equation}

Putting this together, we find
\begin{align}
	\sqrt{-\epsilon_L} &\sim 1- \frac{\lambda}{2}(1-\rat) -\frac{\lambda}{8}(1-\rat)^2 + \cdots \\
	\sqrt{-\epsilon_R} &\sim 1- \frac{\lambda}{2}(1-\rat) -\frac{3\lambda}{8}(1-\rat)^2 + \cdots.
\end{align}
Using \eqref{eq:defr}, we can re-express $\rat$ in terms of the angular velocity $\omega$ of the string,
\begin{equation}
	\omega \sim \frac{2}{1-\rat} + \frac{\lambda}{2}-1 +O(1-\rat),
\end{equation}
to write
\begin{align}
	\epsilon_L \sim -1 +\frac{2\lambda}{\omega} - \frac{\lambda}{\omega^2} + \cdots \\
	\epsilon_R \sim -1 +\frac{2\lambda}{\omega} + \frac{\lambda}{\omega^2} + \cdots.
\end{align}
Finally, translating into energy above the vacuum and angular momentum, this becomes
\begin{equation}\label{eq:specflat}
	E-E_\mathrm{vac} \sim \frac{2\pi}{\omega} T , \qquad J \sim \frac{\pi}{\omega^2} T,
\end{equation}
which is precisely the expected result for a spinning string in flat spacetime without backreaction. 

\subsection{Probe limit ($\lambda \ll 1$)}\label{sec:probe}

Next, we look at the limit of small string tension, holding fixed the size or angular momentum of the string in AdS units. This is the limit in which we expect to recover the results of \cite{Gubser:2002tv} for strings in a fixed AdS background. In terms of our parameters, for this limit we take $\lambda\to 0$ at fixed $\rat$.

In this limit, the string energy will be given by the ground state energy plus a linear correction, so we will have $\epsilon_{L,R}=-1+O(\lambda)$. Using \eqref{eq:defr}, the angular momentum will become  $\omega = \frac{1+\rat}{1-\rat} + O(\lambda)$. We write our expansion in terms of this  physical parameter. In this limit, $x_0$ and $m$ approach the values $x_0 \sim \frac{2}{1+\omega}$, $m\sim \omega^{-2}$.

For the first term in the expressions for $\epsilon_{L,R}$, we use the identity $\Pi(m|m)=\frac{1}{1-m}E(m)$ to find
\begin{equation}
	\tfrac{x_0}{1-\rat}\Pi\left(\left(\tfrac{x_0}{1+\rat}\right)^2\big|m\right) \to \frac{\omega^2}{\omega^2-1} E\left(\omega^{-2}\right).
\end{equation}
The second term is slightly more subtle, requiring the $n\to 1$ expansion
\begin{equation}
	\Pi(n|m) \sim \frac{\pi}{2\sqrt{(1-n)(1-m)}}+K(m) - \frac{1}{1-m}E(m) + O(\sqrt{1-n}), \quad (n\to 1)
\end{equation}
with $1-n=1-(\frac{x_0}{1-\rat})^2 \sim \frac{\lambda^2}{\omega^2-1}$. This gives
\begin{equation}
	\tfrac{x_0}{1+\rat}\Pi\left(\left(\tfrac{x_0}{1-\rat}\right)^2\big|m\right)\sim
	\frac{\pi }{2 \lambda } 
 +\frac{1}{\omega} K\left(\omega^{-2}\right)-\frac{\omega}{\omega^2-1} E\left(\omega^{-2}\right) + O(\lambda^2).
	\end{equation}
Ultimately, we find
\begin{align}
	\sqrt{-\epsilon_L} &\sim  1  + \lambda \frac{2}{\pi}\left(\frac{\omega}{\omega+1}  E\left(\omega^{-2}\right) - K\left(\omega^{-2}\right)\right)+ \cdots \\
	\sqrt{-\epsilon_R} &\sim  1  - \lambda \frac{2}{\pi}\left(\frac{\omega}{\omega-1}  E\left(\omega^{-2}\right) - K\left(\omega^{-2}\right)\right)+ \cdots \;,
\end{align}
which we can write in terms of energy and angular momentum as
\begin{align}
	E-E_\mathrm{vac} &\sim 4T \frac{\omega}{\omega^2-1}  E\left(\omega^{-2}\right) \\
	J &\sim 4T\left(\frac{\omega^2}{\omega^2-1}  E\left(\omega^{-2}\right)- K\left(\omega^{-2}\right)\right).
\end{align}
These are the same results as found in \cite{Gubser:2002tv}, though the methods and integrals we evaluated to get there are entirely different.

In particular, we can now take an additional limit $\omega\to 1$, with $\omega-1 = 2\eta \ll 1$. From this, we find
\begin{equation}
	\Delta \sim \frac{T}{\eta} + T \log \eta^{-1}, \quad J \sim \frac{T}{\eta} - T \log \eta^{-1},
\end{equation}
giving the famous logarithmic anomalous dimension $\Delta-J\sim 2T \log J$. But now this is valid only for $\lambda \ll \eta\ll 1$, in the regime $T\ll J\ll \frac{1}{G_N}$. For sufficiently large spin, we will enter a new regime where gravitational backreaction is important, discussed in section \ref{sec:backreact}.

\subsection{A simple case: $\lambda = 1$}
\label{sec:lambda1}
From the results of the previous section, it appears that there is a special value of the string tension, $\lambda=1$, at which the equations simplify. Specifically, the coefficient of the quartic term in $q(x)$ (defined in \eqref{eq:q}) vanishes, so that it becomes a quadratic and the ratio of the roots $m$ becomes zero. One might expect that this would give rise to some qualitative special feature in the string's behaviour at this value, and perhaps a transition between different regimes for $\lambda<1$ and $\lambda>1$, but this appears not to be the case. 
%Fortunately, this means that the $\lambda=1$ case offers a simple illustration of the qualitative behaviour of
Nonetheless, it happens that the analysis greatly simplifies for this value of the string tension, so this case offers a simple illustration of the qualitative behaviour of the string spectrum at order one values of $\lambda$. 

Since the parameter $m$ of the elliptic functions vanishes in this case, we note their values 
\begin{equation}
\Pi(n|0)= \frac{\pi}{2\sqrt{(1-n)}},\quad K(0)=\frac{\pi}{2}.%, \quad x_0 =\frac{1-\rat^2}{\sqrt{2 \left(\rat^2+1\right)}}.
\end{equation}
Once again we use the angular velocity $\omega$ to parameterise the solutions, and the other quantities are given as
\begin{equation}
	\rat^2 = \frac{\omega^2-2\omega-1}{(1+\omega)^2}, \qquad x_0 = \frac{1}{\omega}+\frac{1}{1+\omega}.
\end{equation}
From this we obtain
\begin{equation}
\label{eq:lambda1sol}
	\sqrt{-\epsilon_L} = \frac{\omega-1}{\omega}, \qquad \sqrt{-\epsilon_R} = \frac{\sqrt{\omega^2-2\omega-1}}{\omega}.
\end{equation}
% Note that in order to avoid closed timelike curves, we need to have positive $\sqrt{-\epsilon_L}$, which means $\omega>1$, thus $\rat^2$ cannot be too negative: $\rat^2>-\frac{1}{2}$. 
% \HM{Comment: $\omega<1$ has CTCs.} 
Re-expressing this in terms of the energy and angular momentum of the string we have
\begin{equation}
	E =\frac{1}{8G_N} \left(\frac{2}{\omega}-1\right), \qquad J = \frac{1}{8G_N \omega^2}.
\end{equation}
Remarkably, these are exactly the same as the flat spacetime results \eqref{eq:specflat}!

However, there is one key difference from the unbackreacted flat spacetime string. In this case, the solution applies only for $\omega>1$, with extrapolation to $\omega<1$ yielding unphysical solutions with closed timelike curves. The family of spinning string solutions terminates at maximal values of energy and angular momentum $E_\mathrm{max}=J_\mathrm{max}=\frac{1}{8G_N}$, corresponding to $h=\frac{c}{8}$, $\bar{h}=\frac{c}{24}$. The relation $E=J>0$ (equivalently $\bar{h}=\frac{c}{24}$, $h>\frac{c}{24}$) corresponds to the energy and angular momentum of an extremal rotating BTZ black hole, and since $E$ is of order $\frac{1}{G_N}$ the corresponding horizon size is of order the AdS length. And indeed, as we take $\omega\to 1$, the solution approaches such a black hole, as we will discuss in more detail in section \ref{sec:extr}.

The special case $\lambda=1$ also admits a closed form solution in terms of elementary functions. In particular, for the radial profile using the $x$ coordinate introduced in \eqref{eq:xcoord} we have the simple result
\begin{equation}\label{eq:xlambda1}
   x(\sigma)=\left(\frac{1}{\omega}+\frac{1}{1+\omega}\right)\cos((\omega-1)\sigma).
\end{equation}

\subsection{Light strings with backreaction}\label{sec:backreact}

Next, we discuss strings with small tension $\lambda\ll 1$, but large angular momentum so that the gravitational backreaction becomes important.

 For the most interesting regime, it turns out that we must tune the parameter $\rat$ so that the quartic polynomial $q$ in \eqref{eq:q} is very close to becoming degenerate (with coincident roots), so $\rat^2$ approaches its minimal value identified in footnote \ref{foot:alphamin}.  This means that the ratio of its roots $m$ is very close to one. It is convenient to use $\mu =1-m$ as an expansion parameter. We must choose how to scale $\mu$ in relation to $\lambda$ as both become very small: we will take $\mu \ll \lambda \ll 1$. In terms of $\rat$, we take
\begin{equation}
	\rat^2 = \frac{\lambda ^2-2+2 \sqrt{1-\lambda ^2}}{\lambda ^2} + \frac{\mu^2}{16}+O(\mu^2\lambda^4) \sim -\frac{\lambda^2}{4},
\end{equation}
and the smaller root of $q$ is given by
\begin{equation}
	x_0^2 =  \frac{2}{1-\lambda ^2+\sqrt{1-\lambda ^2}} -\frac{\mu}{2} + O(\mu\lambda^2) .
\end{equation}

From this, the parameters appearing in the elliptic $\Pi$ functions are
\begin{equation}
	\left(\tfrac{x_0}{1\pm \rat}\right)^2 = 1 \mp i\lambda + O(\lambda^2) + O(\mu).
\end{equation}
We therefore require the expansion
\begin{equation}
	\Pi(n|m) \sim \frac{1}{2(1-n)}\log\left(\frac{4(1-n)}{1-m}\right), \qquad  1-m\ll |1-n| \ll 1,
\end{equation}
where $\log$ is defined with a branch cut along the negative real axis. From this we find
\begin{equation}
	\tfrac{x_0}{1\mp \rat}\Pi\left(\left(\tfrac{x_0}{1\pm\rat}\right)^2\big|m\right) \sim \frac{\pi}{4\lambda} + \tfrac{1}{4}\log\tfrac{1}{\mu} \mp \frac{i}{2\lambda}\log\left(\tfrac{4\lambda}{\mu}\right)+O(1),
\end{equation}
% and the term taking $\rat\to-\rat$ is given by the complex conjugate. 
and (also using the expansion $K(m)\sim \frac{1}{2}\log\left(\frac{16}{1-m}\right)$), we obtain
\begin{equation}
    \sqrt{-\epsilon_L} \sim 1 - \frac{\lambda}{\pi} \log\tfrac{1}{\mu} \,,\qquad 
	\sqrt{-\epsilon_R} \sim \frac{2i}{\pi} \log\left(\tfrac{4\lambda}{\mu}\right), \qquad \text{for } \mu\ll \lambda\ll 1.
\end{equation}
% valid for $\mu\ll \lambda\ll 1$.

We can focus in particular on the interesting regime with twist $\bar{h}$ of order $c$ ($\epsilon_L$ of order unity), which occurs when $\mu$ is exponentially small in $\lambda^{-1}$. Writing
\begin{equation}
	\mu \sim \exp\left(-2\pi\frac{\zeta}{\lambda}\right)
\end{equation}
with $\zeta$ of order one, we have
\begin{equation}
	 \epsilon_L\sim -(1-2\zeta)^2, \quad \epsilon_R\sim \left(\frac{4\zeta}{\lambda}\right)^2,
\end{equation}
or in terms of left- and right-moving conformal dimensions
\begin{equation}
	h \sim \frac{2c}{3}\frac{\zeta^2}{\lambda^2}, \qquad \bar{h} \sim \frac{c}{6}\zeta(1-\zeta) \qquad(0<\zeta<\tfrac{1}{2}).
\end{equation}
The restriction $\zeta<\frac{1}{2}$ is required so that $\sqrt{-\epsilon_L}>0$ as noted at the end of section \ref{sec:strings}. The angular velocity of the spinning string is given by
\begin{equation}
	\omega-1\sim \frac{1-2\zeta}{4\zeta}\lambda^2 \,.
\end{equation}

Now, since this describes a spinning string solution only for $0<\zeta<\frac{1}{2}$, we find that there is a maximal angular momentum $J$ for these solutions,
\begin{equation}
	J_\mathrm{max} \sim \frac{c}{6\pi \lambda^2}\sim \frac{1}{256\pi^3\ell^2 G_N^3 T^2},
\end{equation}
approached in the limit $\zeta\to \frac{1}{2}$. In that limit, the angular velocity $\omega$ approaches unity (from above), and the solution approaches the extremal rotating BTZ black hole. The geometry and the string solution in this interesting limit will be discussed in more detail in section \ref{sec:extr}.

%\begin{equation}
%	\sqrt{\epsilon_R} \sim \frac{2}{\lambda}(1-\sqrt{-\epsilon_L})
%\end{equation}

Note that the limit discussed here (taking $1-m\ll\lambda\ll 1$) does not have overlapping validity with the probe limit discussed in section \ref{sec:probe} (with $m$ of order one). To interpolate between these, one must consider an additional regime where we take $\mu =1-m$ to be of order $\lambda$. In that regime $\epsilon_R$ will be of order unity (spin of order $c$), $\sqrt{-\epsilon_L} =  1 - \tfrac{2}{\pi}\lambda\log{\lambda}^{-1} + O(\lambda)$, and $\omega-1$ is of order $\lambda$. This interpolating regime is technically tricky and not of particular interest, so we do not pursue the details here.

\subsection{The extremal limit for generic $\lambda$}\label{sec:extrgeneric}

Finally, we make some comments on the approach to the extremal limit for values of $\lambda$ of order unity. It turns out that there are two qualitatively different possibilities.

The first possibility occurs for sufficiently small $\lambda <\lambda_c\approx 1.53$, and is illustrated by the simple case $\lambda=1$ described in section \ref{sec:lambda1}, and also by the $\lambda\ll 1$ case in section \ref{sec:backreact}. We have a physically sensible solution for $\rat^2>\rat^2_\mathrm{extr}$, where $\rat^2_\mathrm{extr}<0$ is a critical value corresponding to the extremal solution $\epsilon_L=0$. For $\lambda=1$ we have $\rat^2_\mathrm{extr} = -\frac{1}{2}$ and for $\lambda \ll 1$ we have $\rat_\mathrm{extr}^2 \sim -\frac{\lambda^2}{4}$. When $\rat^2<\rat_\mathrm{extr}^2$, the expression \eqref{eq:rootepsL} for $\sqrt{-\epsilon_L}$ evaluates to a negative value, which gives an unphysical solution as noted at the end of section \ref{sec:strings}. This critical value is otherwise generic, in particular corresponding to a simple zero of $\sqrt{-\epsilon_L}$. This means that $\epsilon_L$ and hence $\frac{c}{24}-\bar{h}$ vanishes quadratically as a function of $h$ at the extremal point. In other words, the Regge trajectories plotted in figure \ref{fig:stringspectrum} are tangent to the black hole threshold line $\bar{h}=\frac{c}{24}$. Also, using \eqref{eq:defr} the same logic tells us that $\omega\to 1$ with $(\omega-1)^2\propto \epsilon_L$. The fact that $\rat_\mathrm{extr}$ is not at any particularly special value also means that there is not a simple general expression for the corresponding value of the spin $J_\mathrm{max}$.

The situation for sufficiently large $\lambda >\lambda_c $ is different. In that case, all $\rat^2<1$ correspond to physically sensible string solutions (since \eqref{eq:rootepsL} gives a positive value for $\sqrt{-\epsilon_L}$), and the extremal limit $\epsilon_L\to 0$ corresponds to taking $\rat^2\to -\infty$. We can see the transition to this behaviour by expanding  \eqref{eq:rootepsL} at $\rat \to i\infty$, finding
\begin{equation}
    \sqrt{-\epsilon_L} \sim \frac{4 (\lambda -1) E\left(-\frac{\lambda +1}{\lambda-1}\right)-4 \lambda  K\left(-\frac{\lambda +1}{\lambda-1 }\right)}{\pi  \lambda\sqrt{\lambda -1}   |\alpha|} \qquad (\alpha\to i\infty).
\end{equation}
The coefficient of $|\alpha|^{-1}$ in this expression is negative for  $1<\lambda< \lambda_c$ and positive for $\lambda> \lambda_c$: its zero defines $\lambda_c$. Since $\sqrt{-\epsilon_L}>0$ for $\alpha=0$, by continuiuty $\sqrt{-\epsilon_L}$ must vanish for some $\alpha_\mathrm{extr}^2<0$ if $\lambda<\lambda_c$ (the first case described above). But for $\lambda>\lambda_c$ there need not be such a zero, and indeed there is not: $\rat_\mathrm{extr}\to i\infty$ as $\lambda\to \lambda_c$. So, in this case $\rat\to i\infty$ is the extremal limit of the spinning string.

From \eqref{eq:rootepsR} we can determine how $h$ behaves in this limit. We have
\begin{equation}
    \sqrt{\epsilon_R} \sim \frac{4 \lambda  }{\pi  \sqrt{\lambda -1}}  \left(K\left(-\tfrac{\lambda +1}{\lambda-1 }\right)-\Pi \left(-\tfrac{1}{\lambda-1}|-\tfrac{\lambda +1}{\lambda-1}\right)\right) + O(\rat^{-2}).
\end{equation}
The constant term here tells us the extremal value of $h$ or $J$. The fact that the second term scales as $\alpha^{-2}$ tells us that as the solution approaches extremality, $h_\mathrm{extr}-h$ is linear in $\epsilon_L$ or in $\frac{c}{24}-\bar{h}$. As a consequence, the Regge trajectories for $\lambda>\lambda_c$ plotted in figure \ref{fig:stringspectrum} are not tangent to the line $\bar{h}=\frac{c}{24}$, but instead meet it with a positive gradient. From \eqref{eq:defr} we can also read off how the angular momentum $\omega$ behaves in the limit: we find it approaches a value $\omega_\mathrm{extr}>1$ which depends on $\lambda$, unlike for $\lambda<\lambda_c$ in which case we always have $\omega\to 1$ in the extremal limit.

% 2. $\lambda<1.53...$: it goes to extremal when $r\to i R$, where $R$ is a finite positive real number. We must have $\omega \to 1$ while $\epsilon_L \to 0$. In this case, $z_L$ goes to some finite value at extremality, whose value depends on $r_*$, the value of $r$ at extremality. Also need to figure out the location of the center...

\section{The approach to an extremal black hole}\label{sec:extr}

A notable feature of the spectra discussed in sections \ref{sec:lambda1} and \ref{sec:backreact} was that the angular momentum and energy of the string could not become arbitrarily large. Instead, our families of spinning string solutions terminated at a finite energy and angular momentum with $\epsilon_L=0$, $\bar{h}=\frac{c}{24}$, or $E=J$. Suggestively, this corresponds precisely to the relation between $E$ and $J$ for an extremal rotating BTZ black hole \cite{Banados:1992gq}. In this section we look at the geometry and string solution in this limit, confirming that the there is indeed a transition to such a black hole.

\subsection{The near-extremal AdS$_2$ region}

Before considering the string solutions, we first examine the geometry in the `near-extremal' limit $\epsilon_R\gg |\epsilon_L|$, concentrating on the case $\epsilon_L<0$ relevant for our spinning strings. 
%For our analysis here, it is more convenient to use $r$ as our radial coordinate defined as 
%\begin{equation}
    %r^2=\frac{(4z+\epsilon_L)(4z+\epsilon_R)}{16z}.
%\end{equation}
% For $\epsilon_L<0$ and $\epsilon_R>0$, $r$ is monotonic with $z$.
For generic values of $r$ of order $\epsilon_R$, which corresponds to $z\sim r^2-\frac{1}{4}\epsilon_R$ of order $\epsilon_R$, the metric \eqref{eq:metricr} approaches that of the extremal rotating BTZ black hole:
\begin{equation}\label{eq:extrBTZ}
    ds^2 \sim -\left(r-\frac{\epsilon_R}{4r}\right)^2 dt^2 + \frac{dr^2}{(r-\frac{\epsilon_R}{4r})^2}+r^2\left(d\phi-\frac{\epsilon_R}{4r^2}dt\right)^2.
\end{equation}
From the double pole in the $dr^2$ term we can see that the geometry develops a `throat' region as $r$ approaches $\frac{\sqrt{\epsilon_R}}{2}$, receding to parametrically large proper distance in the limit. We see the more interesting part of the geometry by zooming in on this region, writing
\begin{equation}
   r=\frac{\sqrt{\epsilon_R}}{2}+\frac{\sqrt{|\epsilon_L|}}{2}\rho
\end{equation}
and keeping $\rho$ fixed in the  limit $ \frac{|\epsilon_L|}{\epsilon_R}\to 0$. In terms of the original $z$ coordinate, this region corresponds to $z$ of order $\sqrt{|\epsilon_L|\epsilon_R}$, with 
\begin{equation}
    \rho \sim \frac{16z^2-|\epsilon_L|\epsilon_R}{8z\sqrt{|\epsilon_L|\epsilon_R}}.
\end{equation}
The scaled metric becomes\footnote{To get this, we should treat $dt$ as order $|\epsilon_L|^{-1/2}$ (accounting for the gravitational redshift in the throat), and $d\phi-dt$ as order unity (as appropriate for corotating observers). This gives a limiting metric that solves the three-dimensional Einstein equations.}
\begin{equation}\label{eq:AdS2}
    ds^2 \sim \frac{1}{4} \left( -4|\epsilon_L| (\rho^2+1)dt^2 +\frac{d\rho^2}{\rho^2+1} \right) +\frac{\epsilon_R}{4} \left( d\phi-dt +2\rho \sqrt{\tfrac{ |\epsilon_L|}{\epsilon_R}} dt \right)^2 \,.
\end{equation}
We recognise the first term as the metric of AdS$_2$ (with two-dimensional AdS length $\frac{\ell}{2}$), while the second term gives a spatial circle of constant radius $\frac{\sqrt{\epsilon_R}}{2}$ fibred over the AdS$_2$ base. This region recedes far from the rest of the geometry: a point with a finite value of $\rho$ is at a parametrically large proper distance (roughly $\frac{1}{4}\log(\frac{\epsilon_R}{|\epsilon_L|})$) from a point with generic $r$ of order $\sqrt{\epsilon_R}$.

Such an AdS$_2$ geometry is familiar from the near-horizon of near-extremal black holes. In that case (corresponding to $\epsilon_L>0$), we obtain a slightly different metric with a horizon of finite (but low) temperature (see section 4 of \cite{Ghosh:2019rcj}, for example). Here there is no horizon to cut off our geometry, which instead terminates at the string worldsheet.

\subsection{Near-extremal strings}

We now examine the location of the string within the near-extremal geometry described above. We focus on determining the innermost and outermost radius of the worldsheet, since this determines whether the string recedes into the long AdS$_2$ throat as it approaches its maximal angular momentum. We examine the small tension $\lambda\ll 1$ regime and the simple $\lambda=1$ example, before commenting on the generic case.

\subsubsection*{Near-extremal strings with $\lambda\ll 1$}

First, consider $\lambda \ll 1$ strings in the regime discussed in section \ref{sec:backreact}, for which $\epsilon_L$ is of order unity (so $0<\bar{h}<\frac{c}{24}$ of order $c$), and $\epsilon_R$ is large (of order $\lambda^{-2}$). %Since $h$ is generic of order $c$ this is perhaps not really approaching a near-extremal black hole, except when we take the additional limit $|\epsilon_L|\ll 1$. Nonetheless, the geometry still has an AdS$_2$ region as in \eqref{eq:AdS2} since we have $\epsilon_R \gg 1$.

To determine the location of the string, we can go back to equations \eqref{eq:z0p}, \eqref{eq:u0p2} and \eqref{eq:v0p2} for $z_0'(\sigma)$, $u_0'(\sigma)$ and $v_0'(\sigma)$ respectively, and simplify in the relevant limit. Using the results in section \ref{sec:backreact}, the parameters $z_{L,R}$ scale as
\begin{equation}
    z_L \sim 2\zeta(1-2\zeta) \lambda^{-2}, \qquad z_R\sim-\tfrac{1}{2}\zeta(1-2\zeta)\sim -\tfrac{1}{4}\lambda^2z_L.
\end{equation}
In particular, $z_L$ determines the $z$ coordinate of the fold of the string, which is also the outermost point (largest $r$) of the string. In terms of the coordinates \eqref{eq:extrBTZ} this is at $r^2\sim\frac{2\zeta}{\lambda^2}\sim \frac{1}{2\zeta}\frac{\epsilon_R}{4}$: since $0<\zeta<\frac{1}{2}$ this is not in the AdS$_2$ region where $r^2\sim\frac{\epsilon_R}{4}$, except after taking an additional limit with $\zeta$ close to $\frac{1}{2}$, when we approach the extremal black hole.

In the AdS$_2$ region where $\rho$ is of order unity, we have $z$ of order $\lambda^{-1}$ given by $z\sim \frac{\zeta(1-2\zeta)}{\lambda}(\rho+\sqrt{1+\rho^2})$. The string equations of motion in this region become
\begin{equation}
    \rho' \sim  2(1-2\zeta)\rho,\qquad
    u_0' \sim  \frac{2\rho}{1+\rho^2},\qquad
    v_0' \sim -\lambda\frac{1-2\zeta}{2\zeta}\frac{1}{1+\rho^2}\,.
\end{equation}
% \begin{equation}
% \begin{aligned}
%     \rho' &\sim  2(1-2\zeta)\rho,\\
%     u_0' &\sim  \frac{2\rho}{1+\rho^2},\\
%     v_0' &\sim -\lambda\frac{1-2\zeta}{2\zeta}\frac{1}{1+\rho^2}\,.
% \end{aligned}
% \end{equation}
It is simple to obtain a solution for $\rho,u_0,v_0$ from these, but for us the main point is to determine the deepest point of the string where $\rho'=0$. This occurs in this AdS$_2$ region at $\rho\approx 0$. Note that the first equation (for $\rho'$) above is valid for $\rho$ of order unity, but not close to the turning point when $\rho$ become sufficiently small.\footnote{In terms of the $x$ coordinate, which scales as $1-x\sim\frac{\lambda}{2}\rho$, for small $\rho$ it becomes important to resolve the nearly-degenerate roots of the quartic $q(x)$ defined in \eqref{eq:q}.} In particular, $u_0$ and $v_0$ both change by an order one amount in that region.

We see that in the extremal limit $\zeta \to \frac{1}{2}$, the outermost point of the string recedes into an AdS$_2$ throat: while we may have $\rho\gg 1$, the ratio between $r$ and the extremal radius $\frac{\sqrt{\epsilon_R}}{2}$ approaches unity so the AdS$_2$ metric \eqref{eq:AdS2} nonetheless applies. The innermost point of the worldsheet resides parametrically deeper still into that throat.

\subsubsection*{Near-extremal strings with $\lambda= 1$}
From section \ref{sec:lambda1}, we can see that the string solution simplifies significantly when we take $\lambda=1$. Here we analyze the near-extremal behaviour, for which $\omega\to 1$ and
\begin{equation}
    \epsilon_L \sim -(\omega-1)^2, \quad \epsilon_R \sim 2-4(\omega-1).
\end{equation}
The extremal angular momentum is $J_\mathrm{max}=\frac{1}{8G_N}$, which means that the horizon radius is of order the AdS scale.

Here, we have an exact solution \eqref{eq:xlambda1} for the radial string profile, with
\begin{equation}
    x(\sigma)\sim \frac{3}{2}\cos((\omega-1)\sigma)
\end{equation}
in the $\omega\to 1$ limit. This coordinate has a simple translation to the AdS$_2$ radial coordinate $\rho$ in this limit, namely
\begin{equation}
    \rho \sim \frac{1-2x^2}{2\sqrt{2}}.
\end{equation}
In particular, values of $x$ of order unity correspond to $\rho$ also of order unity, so the string resides entirely within the AdS$_2$ region, with $-\frac{7}{4\sqrt{2}}\lesssim \rho \lesssim \frac{1}{2\sqrt{2}}$.

% The parameters $z_L$ and $z_R$ scale like
% \begin{equation}
    % z_L \sim \frac{\omega -1}{2}, \quad z_R\sim -\frac{\omega -1}{4}
% \end{equation}

% The outermost point of the string $z=z_L$ in this case is located at $r^2 \sim \frac{1}{2}-\frac{3(\omega-1)}{4}$, and the innermost point is given by $x=x_0 \sim \frac{3}{2}-\frac{5(\omega-1)}{4}$, which is $r^2 \sim \frac{1}{2}-\frac{15(\omega-1)}{8}$. In terms of the $\rho$ coordinate, they both have $\rho \sim \mathcal{O}(1)$, which means they are both located inside the AdS$_2$ throat region.

We can also comment on the solutions for $u_0$, $v_0$ in this example. Expanding the equations of motion to leading order at $\omega\to 1$, we find
\begin{equation}
 u_0' \sim \frac{ (5-\sqrt{2}\rho)}{2(1+\rho^2)} -2,\quad v_0' \sim (\omega-1)\left(\frac{  (2+5\sqrt{2}\rho)}{4(1+\rho^2)}-1\right).
\end{equation}
From the exact radial solution, $\sigma$ varies through the large range $0\leq \sigma \leq \sigma_0= \frac{\pi}{2(\omega-1)}$, so since $u_0,v_0$ must vary over a range of size $\frac{\pi}{2}$ we might expect their derivatives to be of order $\omega-1$. This is true for $v_0$, which is a monotonic function decreasing from $v_0(0)=0$ to $v_0(\sigma_0) = -\frac{\pi}{2}$. But $u_0'$ is of order unity, and $u_0$ decreases to a large value of order $(\omega-1)^{-1}$ before increasing back to $\frac{\pi}{2}$ (to see this order one value requires the next order in the expansion). This might seem strange, but in fact is a result of the fact that the $\tau=0$ slice of our string is highly boosted. To get a better picture of the string we can instead look at its intersection with the $t=0$ slice, which occurs when $\tau = -\frac{1}{2}(u_0(\sigma)+v_0(\sigma))$. This gives
% \begin{equation}
% \begin{gathered}
%      \phi(\sigma) = \frac{\omega+1}{2}v_0(\sigma) -\frac{\omega-1}{2}u_0(\sigma) \\
%      \implies \phi' \sim (\omega-1)\frac{3 \left(2 \sqrt{2} \rho -1\right)}{4 \left(\rho ^2+1\right)} \qquad \text{for }\lambda=1,\   \omega\to 1.
% \end{gathered}
% \end{equation}
\begin{equation}
\begin{gathered}
     \phi(\sigma) = -\frac{\omega+1}{2}v_0(\sigma) -\frac{\omega-1}{2}u_0(\sigma) \\
     \implies \phi' \sim (\omega-1)\left( 2-\frac{7+4\sqrt{2} \rho }{4 \left(\rho ^2+1\right)}\right) \qquad \text{for }\lambda=1,\   \omega\to 1.
\end{gathered}
\end{equation}
From this we see that $\phi'$ is of order $\omega-1$, and always the same sign. In particular, the string winds once around the spatial $\phi$ circle and does not self intersect.

\begin{figure}
    \centering
    \includegraphics[width=.5\linewidth]{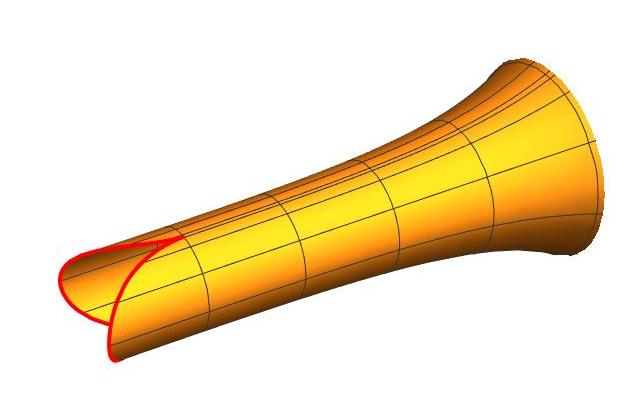}
    \caption{A sketch of the spatial geometry of a $t=0$ slice in the near-extremal  AdS$_2$ region containing the string. The angular direction around the cylinder corresponds to the periodic $\phi$ coordinate, and the direction along the cylinder to the radial $\rho$ coordinate. Going to the right (increasing $\rho$) moves away from the AdS$_2$ region, where the radius of the circle increases: this change of radius becomes important only for $\rho\gg 1$, but has been exaggerated in the figure for illustrative purposes. The geometry is terminated at the left end by the string (marked in red), and the two sides of the string lying on either side of the cusp are identified. Note that this identification does not occur within this slice, identifying points with different values of $t$.}
    \label{fig:AdS2string}
\end{figure}

We have sketched the $t=0$ slice of the AdS$_2$ region of the geometry containing the string in figure \ref{fig:AdS2string}.

\subsubsection*{Near-extremal strings for generic $\lambda$}

We finally comment on the location of the outermost point of the string for generic values of $\lambda$, using the discussion of section \ref{sec:extrgeneric}. Once again, there are two qualitatively different cases depending on whether $\lambda$ is smaller or larger than $\lambda_c$.

For $\lambda<\lambda_c$, in the extremal limit we have $\omega\to 1$ and $\epsilon_L\propto -(\omega-1)^2$, so the $z$-coordinate of the fold of the string scales as $z_L\propto \omega-1$. Since $z_L^2\propto -\epsilon_L$ as both go to zero, this outermost point of the string corresponds to a finite value of $\rho$ which depends on $\lambda$: specifically, $\rho_\mathrm{fold} = \frac{1-|\alpha_\mathrm{extr}|^2}{2|\alpha_\mathrm{extr}|}$. For small $\lambda$ we have $|\alpha_\mathrm{extr}|\ll 1$ so $\rho_\mathrm{fold}\gg 1$, with the string extending close to the boundary of the AdS$_2$ region consistent with our above analysis. As $\lambda$ increases, the string retreats further into the AdS$_2$ region.

As $\lambda\to \lambda_c$, we have $\alpha_\mathrm{extr}\to i\infty$ so $\rho_\mathrm{fold}\to -\infty$, the string retreating far deeper even than the AdS$_2$ region identified above. And indeed, for $\lambda >\lambda_c$, $z_L^2$ goes to zero faster than $\epsilon_L$, which means outer point of the string is much deeper than even the finite $\rho$ AdS$_2$ region.

\subsection{A string to black hole transition?}

It has been suggested before that the states of a single highly excited string may be continuously connected to the internal states of a black hole \cite{Susskind:1993ws,Bowick:1985af,Horowitz:1996nw,Giveon:2006pr,Chen:2021dsw}, a conjecture dubbed the `black hole/string transition'. The fact that our back-reacted spinning string solutions approach the geometry of an extremal BTZ black hole suggests that folded strings in AdS$_3$ may be a novel example of this phenomenon. It is particularly interesting since the resulting black hole has a large horizon area, of order the AdS scale for $\lambda$ of order unity and larger still for small $\lambda$.% This idea deserves further scrutiny; here we make a few small comments.

We have studied only classical solutions of the gravitating string, but when the solution is sufficiently close to the extremal limit we expect quantum effects to become important. Such effects are now well-understood for the low-temperature limit of BTZ black holes approaching the extremal limit from energies above the threshold (rather than below as for the strings). For energies of order $G_N$ above the threshold a perturbative gravitational mode describing fluctuations of the long AdS$_2$ throat becomes strongly coupled \cite{Ghosh:2019rcj}, but nonetheless remains under good control since it is described by the solvable Schwarzian theory \cite{Jensen:2016pah,Maldacena:2016upp,Engelsoy:2016xyb}, similarly to near-extremal black holes in higher dimensions. For exponentially small temperatures (when there are only order one available  microstates) fluctuations of topology become important \cite{Saad:2019lba,Maxfield:2020ale} and the gravitational description remains a mystery; stringy physics may play an significant role here. We expect a similar Schwarzian mode to become important for the near-extremal spinning string, as well as quantum fluctuations of the string. Perhaps by including such effects it is possible to interpolate between classical spinning strings and classical BTZ black holes through a near-extremal quantum regime.

\section{Discussion}
\label{sec:discussion}

\subsection{Spinning strings in top-down models}\label{sec:D1D5}

We studied the simplest possible `bottom-up' model of a Nambu-Goto string coupled to Einstein gravity. How can we embed this physics in complete top-down string theory constructions?

We'll take the first steps towards this by examining the parameters in the paradigmatic example of strings in AdS$_3$, the D1-D5 system. This arises from the near-horizon description of $Q_1$ D1-branes and $Q_5$ D5-branes wrapped on a four-manifold $\mathcal{M}_4$ (either $T^4$ or K3) in type IIB string theory. The resulting geometry is AdS$_3\times S^3\times \mathcal{M}_4$, with the radius of $S^3$ equal to the AdS length $\ell$. The AdS scale and three-dimensional Newton's constant $G_N$ are given by \cite{Aharony:1999ti}
\begin{equation}
    \ell^2 = g_6 \sqrt{Q_1 Q_5} \ell_s, \quad G_N = \frac{g_6^2\ell_s^4}{4\ell^3},
\end{equation}
where $g_6$ is the string coupling in six dimensions (after compactifying on $\mathcal{M}_4$). This means that our parameter $\lambda$ determining the strength with which fundamental string couples to gravity scales as
\begin{equation}
    \lambda \propto \frac{g_6}{\sqrt{c}}
\end{equation}
For weakly coupled strings ($g_6\ll 1$) and large-radius AdS in Planck units ($c\propto Q_1 Q_5\gg 1$), we have $\lambda \ll 1$, so our analysis of that case is most relevant for  this top-down model. In particular, the maximal angular momentum where the string merges with black holes for this case would be at $J_\mathrm{max}$ of order $g_6^{-2}$.

% From this, it appears that the largest possible value of $\lambda$ is of order unity, though this would occur only for examples with small central charge in the strongly-coupled region of moduli space where it is unclear whether our classical gravitational analysis has any relevance.% Larger values of $\lambda$ could be relevant in the usual classical regime for other strings in the theory (D-strings and D5 branes wrapped on $\mathcal{M}_4$), though we have not examined parameters for those strings.

Importantly, in this D1-D5 example there is no background NS-NS $B$-field, which would be important to include since it couples directly to the fundamental string. See section \ref{sec:Bfield} for comments on the inclusion of NS-NS flux.  For a similar reason, our results do not apply to the D-string since the D1-D5 background is supported by RR two-form flux, which couples to the D1 in the same way that the NS field coupled to the fundamental string.

As mentioned above, the geometry for the D1-D5 system is not simply AdS$_3$, but includes a compact manifold $S^3\times \mathcal{M}_4$ (and the size of the $S^3$ is always the same as the AdS scale). Since our results do not make reference to this internal space, they apply when the string does not have significant momentum in the compact directions, and when back-reaction does not significantly alter the internal geometry. In particular this applies in the sector of zero R-charge, for which the wavefunction of the string is independent of the compact directions. It would be interesting to generalise to find back-reacted solutions for strings which are localised and carry momentum in the internal directions.

\subsection{The CFT$_2$ dual of spinning strings}

Perhaps the most obvious question is whether we can understand the states dual to spinning strings in a holographic dual conformal field theory. Two complementary approaches to this question spring to mind: one might attempt to identify these states in a generic theory with minimal assumptions following a bootstrap philosophy, or alternatively to compute their spectrum in a specific top-down holographic model.

\subsubsection*{Bootstrap}

This work was originally motivated by results using analytic bootstrap methods to constrain the spectrum of irrational two dimensional CFTs. These results give a gravitational characterisation of the large spin states of generic theories, but show no sign of strings. What role (if any) do string states play in this context, and how might they be identified and constrained by the bootstrap?

Specifically, consistency demands \cite{Collier:2018exn,Kusuki:2018wpa,Collier:2019weq} that a generic\footnote{More precisely, the results in question apply for a unitary theory with $c>1$ and a Virasoro twist gap: the conformal dimensions $h$ of Virasoro primary states except the vacuum have a positive lower bound. This is roughly the requirement that the theory has no conserved currents besides Virasoro descendants of the vacuum (though technically slightly stronger).} irrational CFT$_2$ contains states corresponding to both multi-particle states in AdS and to BTZ black holes. These exist in a large spin limit of fixed $\bar{h}$ and $h\to\infty$, shown by bootstrapping four-point functions, in which states with the spectrum suggested by their AdS analogues must appear as intermediate states to satisfy crossing symmetry. The two-particle states are CFT$_2$ versions of double-twist operators in higher dimensions \cite{Fitzpatrick:2012yx,Komargodski:2012ek}, with the novelty in $d=2$ that their spectrum is modified (in a universal manner controlled by the central charge $c$ only) by gravitational interactions. Black holes do not appear in the analogous higher dimensional bootstrap, but for CFT$_2$ are necessary for modular invariance \cite{Afkhami-Jeddi:2017idc,Collier:2016cls} as well as four-point crossing. Corrections to this spectrum away from the large spin limit also have gravitational descriptions in generic theories. This includes anomalous dimensions of two-particle states due to interactions from the four-point function bootstrap \cite{Collier:2018exn}, as well as quantum corrections to the BTZ black hole threshold by combining the existence of multi-twists with modular invariance \cite{Maxfield:2019hdt}. Finally, theories with weakly coupled local holographic duals are precisely those for which the corrections to these bootstrap results are small, which means that they can remain true for generic kinematics and not only at large spin.

Since the CFT$_2$ bootstrap has proven so powerful at recovering the spectrum of AdS$_3$ gravity, it is natural to ask why string states have not yet appeared from this perspective, and whether their spectrum can be similarly determined or constrained. Our results indicate that they are unlikely to be visible from the conventional large-spin (lightcone) bootstrap, since the strings have a maximal $J$ before merging with the BTZ spectrum, which explains why they were not apparent in the previous work reviewed above. Nonetheless, perhaps a combination of ideas from large $c$ and large $J$ analyses will give access to these states. To this end, it may be helpful to understand whether the spinning strings can ever be the dominant intermediate states in a correlation function corresponding to high-energy bulk scattering of particles, and if so in what kinematic regime.

\subsubsection*{In top-down constructions}

A complementary approach to the above is to study the spectrum in a specific realisation of AdS$_3$ gravity with a known CFT dual. For example, in the D1-D5 system the dual CFT can be described as a deformation of a free symmetric orbifold theory (and likewise the recently proposed theories in \cite{Belin:2020nmp}). The free orbifold point corresponds to $\lambda=0$, where our leading Regge trajectory of string states becomes a tower of higher spin single-trace conserved currents (with $\bar{h}=0$). From this we expect that no string back-reaction effect remains at strictly zero coupling. But under deformation away from this point these current acquire anomalous dimensions, which we hope to relate to the classical string spectrum including back-reaction. A conformal perturbation theory analysis is indicative of the expected logarithmic spin dependence from the GKP strings \cite{Gaberdiel:2015uca}; our results indicate that this will break down at sufficiently large spin (when $J$ is of order the inverse square of the coupling).

Since the interesting regime of large spin involves gravitational back-reaction, there is a sense in which the string states of interest are multi-particle states of many gravitons (along with the string itself). This intuition suggests that from the CFT, we will have significant mixing between multi-trace operators consisting of the long string states dressed with many stress tensors. If this is correct, taking the mixing into account may make the problem significantly more technically challenging.

\subsection{Background $B$-field}\label{sec:Bfield}

Besides the importance of gravitational back-reaction, there is another well-known way in which AdS$_3$ is special: it can support a background NS-NS $B$-field which respects the symmetries. This is a two-form potential with gauge-invariant three-form field strength $H=dB$; a nonzero value can respect the symmetries of AdS$_3$ since we can choose $H$ to be proportional to the volume form. It couples to the string by a term $\int B$ in the action (integrating the pullback of $B$ on the worldsheet). For this reason, our results need to be modified in a background with this field turned on. Previous results (without back-reaction) include \cite{David:2014qta,Banerjee:2015qeq}.

To gain some intuition, it is helpful to use Stokes' theorem to write the coupling of $B$ to the string as the integral of $H$ over a three-dimensional region bounded by the string worldsheet. This action becomes proportional to the (signed) volume contained within the string, which we can think of as providing a `pressure' force supporting a bubble bounded by the worldsheet, while the usual Nambu-Goto action is analogous to a surface tension for this bubble. From this it is clear that our folded strings will no longer be classical solutions: the pressure force will cause them to `puff up', separating the two strands of folded string. This means that the method we used to find solutions does not straightforwardly carry over to a case when this field is nonzero. On the other hand, in this case there are simpler static solutions with rotational symmetry --- circular strings held in equilibrium by the balance of `surface tension' and `pressure' forces --- and it may be interesting to analyse these including back-reaction.

We note that there is a special value of NS-NS flux in AdS$_3$. Since the area contained in a large circle in the hyperbolic plane is proportional to its perimeter, the forces on a string as it approaches the AdS conformal boundary can be balanced by tuning the $B$-field strength. This is the `pure NS point', an example of which is the S-dual of the D1-D5 system above (for which the D1s become fundamental strings, and D5s become NS5s), with the geometry supported by NS-NS flux only instead of R-R flux only. In this case, the worldsheet theory becomes an $SL(2,\mathbb{R})$ WZW model, which makes construction of classical solutions much simpler \cite{Giveon:1998ns,Maldacena:2000hw,Loewy:2002gf}. Perhaps some of this simplicity will remain once back-reaction is taken into account.

\subsection{Excitations}

The spinning strings we constructed correspond to the ground state of the string for given angular momentum. It is clearly of interest to study more general classical solutions, or perhaps the quantum theory of excitations on top of our solutions. Our methods relied heavily on the symmetry of the spinning string, which will be broken by more general states, so a new idea is required. For quantization of the GKP string, see \cite{Frolov:2002av,Tseytlin:2002ny}.

It is of particular interest to count the excited states of the string in the near-extremal limit as a probe of the black hole/string transition. Perhaps the entropy of string states smoothly crosses over through the transition to the Bekenstein-Hawking entropy of near-extremal rotating black holes?

\acknowledgments
We would like to thank Don Marolf for providing useful comments on our manuscript. HM is supported by DOE grant DE-SC0021085 and a Bloch fellowship from Q-FARM. HM was also supported by NSF grant PH-1801805, by a DeBenedictis Postdoctoral Fellowship,  and by funds from the University of California. ZW was supported by NSF grants PHY-1801805 and PHY-2107939, and by funds from the University of California.

\cleardoublepage
\addcontentsline{toc}{section}{References} 
\bibliographystyle{JHEP} 
\bibliography{StringsAdS3.bib}

\end{document}